%% file: conference_101719.tex
\documentclass[conference]{IEEEtran}
\IEEEoverridecommandlockouts
\usepackage{cite}
\usepackage{amsmath,amssymb,amsfonts}
\usepackage{algorithmic, algorithm}
\usepackage{graphicx}
\usepackage{textcomp}
\usepackage{xcolor}
\usepackage{listings}
\usepackage{pgfplots}
\usepackage{pgfplotstable}
\usepackage{tikz-timing}
\usepgfplotslibrary{groupplots}
\usepackage{multirow}
\usepackage[hidelinks]{hyperref}
\usepackage{soul}
\usepackage{orcidlink}
\def\BibTeX{{\rm B\kern-.05em{\sc i\kern-.025em b}\kern-.08em
    T\kern-.1667em\lower.7ex\hbox{E}\kern-.125emX}}

\newcommand*\circled[1]{\tikz[baseline=(char.base)]{
            \node[shape=circle,fill,inner sep=1pt] (char) {\textcolor{white}{#1}};}}

\pgfplotsset{compat=1.18}

\begin{document}

\title{Uncertainty-Guided Live Measurement Sequencing for Fast SAR ADC Linearity Testing}

\author{\IEEEauthorblockN{Thorben Schey\IEEEauthorrefmark{1}\orcidlink{0009-0003-0775-8762},
Khaled Karoonlatifi\IEEEauthorrefmark{2}\orcidlink{0009-0004-7135-6749},
Michael Weyrich\IEEEauthorrefmark{1}\orcidlink{0000-0003-3176-9288}
and Andrey Morozov\IEEEauthorrefmark{1}\orcidlink{0000-0001-6772-8889}}\\
\IEEEauthorblockA{\IEEEauthorrefmark{1}Institute of Industrial Automation and Software Engineering, University of Stuttgart\\
Stuttgart, Germany\\
Email: \{first.last\}@ias.uni-stuttgart.de\\
\IEEEauthorrefmark{2}Advantest Europe GmbH\\
Böblingen, Germany\\
Email: \{first.last\}@advantest.com}}
\maketitle

\IEEEpubid{\begin{minipage}{2.0\columnwidth}
\vspace{50pt}
\centering\footnotesize
© 2025 IEEE. Personal use of this material is permitted. Permission from IEEE must be obtained for all other uses, including reprinting/republishing for advertising or promotional purposes, creating new collective works, for resale or redistribution to servers or lists, or reuse of any copyrighted component of this work in other works.
\end{minipage}}

\begin{abstract}
This paper introduces a novel closed-loop testing methodology for efficient linearity testing of high-resolution Successive Approximation Register (SAR) Analog-to-Digital Converters (ADCs).
Existing test strategies, including histogram-based approaches, sine wave testing, and model-driven reconstruction, often rely on dense data acquisition followed by offline post-processing, which increases overall test time and complexity.
To overcome these limitations, we propose an adaptive approach that utilizes an iterative behavioral model refined by an Extended Kalman Filter (EKF) in real time, enabling direct estimation of capacitor mismatch parameters that determine INL behavior.
Our algorithm dynamically selects measurement points based on current model uncertainty, maximizing information gain with respect to parameter confidence and narrowing sampling intervals as estimation progresses.
By providing immediate feedback and adaptive targeting, the proposed method eliminates the need for large-scale data collection and post-measurement analysis.
Experimental results demonstrate substantial reductions in total test time and computational overhead, highlighting the method’s suitability for integration in production environments.
\end{abstract}


\section{Introduction}
\label{sec:Introduction}
Analog-to-Digital Converters (ADCs) are essential components in modern electronic systems, serving as interface elements between analog and digital circuitry.
To ensure reliable operation, comprehensive testing of their key performance parameters is essential.
In particular, linearity testing, which is characterized by Differential Non-Linearity (DNL) and Integral Non-Linearity (INL), is crucial as it directly affects the accuracy and practical performance of the ADC.

Traditionally, the histogram test is widely adopted to determine the DNL and INL of ADCs by analyzing large volumes of sampled data, typically involving extensive measurement sets to achieve sufficient statistical accuracy.
While this approach is effective and straightforward to implement, it is time-consuming and requires the transfer of large amounts of data to test workstations.

In production environments, minimizing test duration and data transfer is increasingly important - especially for ADC architectures like the Successive Approximation Register (SAR), which are widely used in power- and area-constrained applications.
Their binary-weighted internal structure gives rise to predictable, architecture-specific nonlinearity patterns, which can be exploited to guide the measurement process.

Recent research has introduced methodologies utilizing mathematical models and segmented testing approaches to reduce the number of necessary measurements drastically \cite{chaganti2018low, yu2012algorithm}.
While these techniques significantly improve test efficiency, they typically rely on full data acquisition followed by offline parameter estimation or optimization procedures.
This post-processing overhead limits their practicality in time-critical or resource-constrained production settings \cite{aswin2023new, chaganti2018fast}.
Until now, there has been no approach that performs dynamic measurement selection during test execution or integrates real-time model refinement into the acquisition process.
This opens the door for new strategies that adaptively focus measurements where uncertainty is greatest, enabling efficient closed-loop estimation without the need for external computation.

\textbf{Contributions:}
This work introduces a novel closed-loop testing methodology for high-resolution SAR ADCs, embedding the device under test into an iterative framework that adaptively refines a behavioral mismatch model in real time.
A residual-based adaptive Extended Kalman Filter (EKF) is used to continuously update parameter estimates from incremental measurements, with each code edge dynamically chosen to maximize model refinement.
Unlike existing approaches relying on bulk data collection and post-processing, the proposed method performs all estimation steps during the test sequence, eliminating the need for offline computation and large-scale data transfer.
Only the compact model parameters need to be returned, enabling minimal bandwidth use.
Measurement points are selected dynamically based on current model uncertainty, aiming to maximize information gain per measurement and minimize redundant sampling.
This strategy reduces overall test time and supports efficient in-situ characterization, making it well suited for fast, resource-aware test procedures.

The remainder of this paper is organized as follows:
\autoref{sec:State_of_the_Art} reviews ADC error metrics, the SAR ADC architecture, existing test strategies, and introduces the standard EKF formulation.
In \autoref{sec:Proposed_Method} our proposed testing strategy is presented.
\autoref{sec:Experimental_Results} shows experimental results.
\autoref{sec:Discussion} analyzes the findings, and finally \autoref{sec:Conclusion} summarizes key benefits and future directions.

\section{State of the Art}
\label{sec:State_of_the_Art}
\subsection{ADC behavior and error metrics}
\label{secsub:State_of_the_Art_ADC_Metrics}
Analog-to-Digital Converters (ADCs) discretize a continuous analog input within a defined range into one of $2^N$ digital codes, where $N$ is the resolution.
The input voltage range is bounded by $V_{\mathrm{ref}-}$ and $V_{\mathrm{ref}+}$, forming the Full-Scale Range:

\begin{equation}
    \label{eq:FSR}
    \mathrm{FSR} = V_{\mathrm{ref}+} - V_{\mathrm{ref}-}
\end{equation}

This range is divided into Full Scale $\mathrm{FS} = 2^N$ codes, resulting in $\mathrm{FS} - 1$ transition points, or code edges, between adjacent codes.
The size of one quantization step - referred to as the Least Significant Bit (LSB) - is given by:

\begin{equation}
    \label{eq:LSB}
    \mathrm{LSB} = \frac{\mathrm{FSR}}{\mathrm{FS}}
\end{equation}

The key static linearity metrics in specification-based ADC testing are Differential Non-Linearity (DNL) and Integral Non-Linearity (INL), both expressed in LSB. They quantify how closely the ADC behavior matches its ideal code edge transfer curve.
The DNL at code $c$ is thereby defined as the deviation in the length of its ideal quantization interval of $1\,\mathrm{LSB}$:

\begin{equation}
    \label{eq:DNL}
    \mathrm{DNL}[c] = \mathrm{CE}[c+1] - \mathrm{CE}[c] - 1
\end{equation}
with $\mathrm{CE}[c]$ denoting the edge between $c$ and $c+1$, in LSB.

The INL measures the deviation of each code edge from its ideal position on the code edge transfer curve.
It is given by:
\begin{equation}
    \label{eq:INL}
    \mathrm{INL}[c] = \mathrm{CE}[c] - \mathrm{CE}_{\mathrm{ideal}}[c]
\end{equation}

Offset and gain errors, as well as missing codes, are additional static metrics evaluated during ADC testing.
While missing codes directly affect DNL and INL, offset and gain errors are typically removed before linearity analysis and are therefore not captured in the reported DNL or INL values.
During production testing, the maximum absolute values of DNL and INL are compared against predefined limits to determine whether a device passes specification \cite{rapuano2005adc, vasan2011adc}.

\subsection{Successive Approximation Register ADCs}
\label{secsub:State_of_the_Art_SAR}
SAR ADCs are widely used in applications requiring moderate speed, low power, and small area, such as portable electronics and sensor interfaces \cite{rapuano2005adc, vasan2010linearity}.
Compared to flash or pipelined ADCs, they offer a favorable trade-off between resolution, power consumption, and complexity.
The core of a SAR ADC consists of a sample-and-hold stage, a binary-weighted Digital-to-Analog Converter (DAC), a comparator, and the successive approximation logic (see \autoref{fig:SAR_ADC_Schematic}).
During conversion, the input is sampled and the DAC conducts a binary search, guided by comparator decisions comparing the DAC output voltage to the input.
Using this feedback, the SAR logic updates the DAC input for the next bit decision \cite{michaeli2008unified}.

The ADC's resolution is set by the number of successive comparisons, requiring $N$ clock cycles and resulting in predictable timing and power consumption.
Given their architecture, SAR ADCs have certain shortcomings - some relate to static linearity, others to dynamic behavior or long-term drift.
Common error sources include:

\begin{figure}[t]
    \setlength{\belowcaptionskip}{-5pt} 
    \centering
    \includegraphics[width=\linewidth]{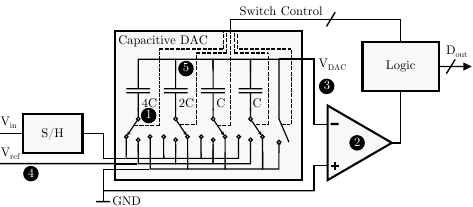}
    \caption{Schematic of a SAR ADC including a capacitive DAC with binary-weighted structure.
    Core components such as the sample-and-hold input stage, DAC switch array, comparator, and control logic are shown.}
    \label{fig:SAR_ADC_Schematic}
\end{figure}

\begin{itemize}
    \item[\circled{1}] \textbf{DAC capacitor mismatch:} Variations in the unit or binary-weighted capacitors within the DAC lead to errors in the generated reference voltages during the approximation steps.
    This is one of the dominant sources of DNL and INL errors in SAR ADCs \cite{wegener2000model, michaeli2008unified}.
    
    \item[\circled{2}] \textbf{Comparator offset and noise:} A static offset in the comparator shifts the decision thresholds, while noise introduces variability in the bit decision.
    These effects can directly propagate into the final digital output and impact DNL and INL \cite{huang2016analysis}.
    
    \item[\circled{3}] \textbf{DAC settling errors:} Incomplete settling of the DAC voltage between comparison steps results in incorrect comparator input, especially at higher sampling rates.
    While this primarily affects dynamic behavior, it may also introduce systematic deviations in code edge locations if the settling error is signal-dependent \cite{vasan2010linearity}.
    
    \item[\circled{4}] \textbf{Reference voltage instability:} Supply ripple or reference voltage droop can cause variation in the DAC output and comparator threshold.
    These effects typically degrade dynamic performance but may contribute to INL if the instability follows a spatial or temporal pattern \cite{huang2016analysis}.
    
    \item[\circled{5}] \textbf{Parasitic capacitance and crosstalk:} Layout-induced parasitics can cause coupling between DAC nodes or between digital and sensitive analog paths.
    These nonlinear effects are difficult to model but may appear as subtle shifts in DNL and INL \cite{zhao2022evaluation}.
\end{itemize}

Among these, capacitor mismatch and comparator offset are the dominant contributors to static nonlinearity and are the focus of most INL/DNL modeling efforts \cite{aswin2023new, brenna2015tool}.
Other effects, such as DAC settling and reference ripple, are harder to capture in static models and not considered in this work.

The bit-wise switching sequence in SAR ADCs leads to characteristic INL/DNL patterns, which have been exploited in test strategies to reduce measurement effort \cite{chen2015ultrafast, chaganti2018fast, aswin2023new}.

\subsection{Existing ADC test strategies}
\label{secsub:State_of_the_Art_TestStrategies}
Numerous methods exist for evaluating static ADC nonlinearity, particularly DNL and INL.
The most common approach in production test environments is the ramp-based histogram test.
A highly linear ramp is applied to the input, and a histogram of output codes is used to estimate DNL, while INL is derived by summing the individual DNL values.
It typically needs 20-100 Hits Per Code (HPC), which for a 16-bit ADC is about 1.3-6.5 million samples, making it impractical for high-resolution ADCs without long acquisition times \cite{goyal2005test,vasan2011adc,fu2024stimulus}.

An alternative is the sine wave histogram test.
Instead of requiring a perfectly linear ramp, this method uses a spectrally pure sine input and leverages its known probability density function to reconstruct code densities.
While this relaxes input linearity requirements, it leads to uneven sampling across the code space, with far fewer hits at the center of the ADC range - precisely where many SAR ADCs exhibit their most critical nonlinearity \cite{vasan2011adc,zhao2022evaluation}.

To improve test time, recent advances such as the code-selective histogram method have been developed.
Zhao et al. proposed using a two-tone sine wave input to concentrate sampling on specific vulnerable codes, particularly those triggered by MSB transitions.
It improves local accuracy with fewer samples, balancing test time and measurement precision \cite{zhao2022evaluation}.

Servo-loop methods embed the ADC in a feedback loop with a DAC to directly and accurately measure code edge locations, avoiding statistical noise \cite{corcoran1975high, wegener2000model}.
However, each edge measurement requires loop settling time - typically on the order of 20-40 ms - which can limit scalability. 
Therefore, these methods are suited for sparse, high-precision verification rather than full-scale characterization \cite{wegener2000model, rapuano2005adc}.

More recently, model-based techniques have drastically improved test efficiency.
An innovative contribution came from Yu and Chen \cite{yu2012algorithm}, who proposed modeling INL using a segmented nonparametric approach, treating the transfer function as composed of locally correlated error segments.
This insight was further developed into the Ultrafast Segmented Model Identification of Linearity Errors (uSMILE) method by Chaganti et al. \cite{chaganti2018fast}.
uSMILE significantly reduces required data acquisition by replacing histogram-based code-by-code estimation with parametric model fitting across MSB-, ISB-, and LSB-level segments.
In 16-bit ADCs, sample counts can drop from 4 million to under 40,000 without loss in accuracy.

Later variants of uSMILE, including runtime-optimized and extrapolation-based approaches \cite{chaganti2018fast, aswin2023new}, further reduce memory and processing overhead by reconstructing INL from sparse measurement grids.
These methods are particularly effective for SAR ADCs, whose binary-weighted DACs produce structured nonlinearity patterns.

Importantly, the benefits of reduced sampling time come at the cost of more involved post-processing.
uSMILE and its variants require solving matrix equations, segment parameter estimation, and optionally applying regularization.
In production test environments, this adds a computational load that must be carefully managed depending on whether processing occurs on-chip, at the tester, or after data transfer.

Li et al. \cite{li2024low} presented a segmented polynomial model that combines structured and unstructured nonlinearity sources.
It supports piecewise modeling of INL with nonuniform segment lengths and polynomial degrees, further improving model fitting quality for architectures such as SAR ADCs where nonlinearity is not evenly distributed across the input range.

Another research direction has aimed to improve post-processing efficiency and robustness using Kalman filtering techniques.
Jin et al. \cite{jin2006linearity} demonstrated that INL behavior can be treated as a state estimation problem, where successive code observations refine an internal model of the ADC’s code edge transfer curve.
Kalman filtering allows recursive updating of model parameters and noise filtering without needing dense sample coverage.
Vasan and Chen \cite{vasan2010linearity} extended this idea in SEIR-KF and KHK-KF methods, combining stimulus error identification with optimal estimation to reduce sampling requirements.
These approaches enable INL estimation with fewer HPCs - e.g., reducing from 4 HPC to 1 HPC - while retaining the same level of accuracy.
However, they are typically applied as post-processing steps to histogram data, and thus do not remove the need for full-data acquisition unless combined with stimulus-aware strategies.

In summary, conventional ramp and sine histogram methods are reliable but relatively slow, due to their high sampling needs and basic statistical nature.
Servo-loop methods are precise but require substantial time for loop settling at each code edge.
uSMILE and its successors represent the current state of the art, offering massive reductions in sampling time through model-based estimation, while requiring moderate post-processing.
Kalman filter approaches serve as powerful complements to these techniques, improving accuracy and robustness during reconstruction.
In all cases, total test time includes both data acquisition and post-processing, both critical in production planning.

\subsection{Extended Kalman Filter for static parameter estimation}
\label{secsub:State_of_the_Art_EKF}
The Extended Kalman Filter (EKF) is a widely used approach for recursive state estimation in systems with nonlinear measurement functions.
It extends the standard Kalman filter by linearizing the observation model around the current estimate using a Jacobian \cite{simon2006optimal}.
In the general EKF formulation, the system evolves as:

\vspace{-6pt}
\begin{align}
    \boldsymbol{\theta}_{k+1} &= \boldsymbol{\theta}_k + \boldsymbol{w}_k, \\
    \mathbf{z}_k &= h(\boldsymbol{\theta}_k) + \boldsymbol{\nu}_k,
\end{align}

where $\boldsymbol{\theta}_k$ is the parameter vector at iteration $k$, $\boldsymbol{w}_k$ the process noise, $\mathbf{z}_k$ the measurement, $h(\cdot)$ the nonlinear measurement function, and $\boldsymbol{\nu}_k$ the measurement noise.
To apply the Kalman filter update, this function is linearized around the current estimate $\boldsymbol{\theta}_k$ using its Jacobian matrix:

\begin{equation}
    \mathbf{H}_k = \left. \frac{\partial h(\boldsymbol{\theta})}{\partial \boldsymbol{\theta}} \right|_{\boldsymbol{\theta}_k}
\end{equation}

$\mathbf{H}_k$ is then used in place of the observation matrix in the Kalman gain and update equations, enabling the filter to approximate the nonlinear model with a local linear one.

In static estimation problems, such as the one considered here, the system parameters do not evolve over time.
The process noise \( \boldsymbol{w}_k \) is therefore set to zero, and the estimation is driven solely by incoming measurements.
This simplifies the recursion while preserving the ability to quantify and adapt uncertainty.

The EKF was chosen for this work due to its favorable balance of accuracy and computational simplicity.
While other nonlinear estimation techniques, such as the Unscented Kalman Filter (UKF), Particle Filter, or Moving Horizon Estimation (MHE), offer improved modeling capabilities in highly nonlinear or dynamic systems, they are computationally more demanding.
In contrast, the mismatch model used here is static and only moderately nonlinear, allowing accurate local linearization and stable filter behavior.
The EKF supports recursive updates with low overhead, making it well suited for integration into real-time test loops without requiring external optimization or post-processing.

\section{Proposed Method}
\label{sec:Proposed_Method}

\subsection{Conceptual overview}
\label{secsub:ProposedMethod_ConceptOverview}
The proposed Uncertainty-Guided Live Measurement Sequencing (UGLMS) method applies a modified EKF to estimate capacitor mismatch parameters in SAR ADCs through an iterative, closed-loop test process.
Instead of sampling the entire transfer curve, the algorithm targets only a small number of carefully selected code edges and uses localized voltage sweeps to infer deviations from ideal behavior.

At its core, the method maintains an internal estimate of the mismatch vector, which is iteratively refined.
In each iteration, the code edge with the highest expected information gain is selected for probing.
Around that code edge, a localized voltage sweep is applied using a high-resolution DAC, and the resulting ADC output samples are aggregated to compute a scalar measurement.
This scalar measurement represents the observed deviation between expected and actual code edge positions and is used to update the mismatch estimate through a specialized EKF update step.

The filter operates in a static estimation setting, where the parameters remain constant and are updated solely based on measurement feedback.
Its recursive structure supports real-time convergence during the test process and eliminates the need for post-processing.
To improve robustness under quantization and input variability, a residual-based covariance inflation mechanism is integrated into the update loop.

\subsection{Filter structure and simplifications}
\label{secsub:ProposedMethod_FilterStructure}
The EKF structure applied in this work estimates a static parameter vector $\boldsymbol{\theta} \in \mathbb{R}^N$, representing capacitor mismatches in a SAR ADC with $N$ bits.
The measurement model is defined as a nonlinear function $f_c(\boldsymbol{\theta})$, which predicts the code edge voltage at a given code $c \in \{0, \ldots, 2^N-1\}$.

In each iteration $k$, a scalar measurement $z_k$ is obtained by applying a localized voltage sweep around the selected code edge and averaging the resulting ADC output samples.
This measurement is modeled as the deviation between predicted and observed transition behavior:

\begin{equation}
    z_k = f_c(\boldsymbol{\theta}) - f_c(\hat{\boldsymbol{\theta}}_k) + \nu_k,
\end{equation}

where $\hat{\boldsymbol{\theta}}_k$ is the current mismatch estimate, and the measurement noise $\nu_k$ is modeled as zero-mean Gaussian with variance $R$, i.e., $\nu_k \sim \mathcal{N}(0, R)$.

Since the ADC model is static and the functional dependence $f_c(\boldsymbol{\theta})$ remains unchanged during estimation, the full Jacobian matrix $\mathbf{H} \in \mathbb{R}^{C \times N}$ can be precomputed once before the first iteration, where each row corresponds to a code $c$.
During each update, only the row corresponding to the currently evaluated code is retrieved:

\begin{equation}
    \mathbf{j}_c = \left. \frac{\partial f_c(\boldsymbol{\theta})}{\partial \boldsymbol{\theta}} \right|_{\hat{\boldsymbol{\theta}}_k}.
\end{equation}

This localized retrieval avoids unnecessary computations and remains consistent with the measurement strategy: since each scalar observation provides information about only one code edge, using additional Jacobian rows from unmeasured codes would introduce unrelated and unjustified updates.

With this simplified structure, the EKF update equations for the scalar-output case become:

\begin{align}
    \mathbf{K}_k &= \frac{\mathbf{P}_k\,\mathbf{j}_{c}^\top}{\mathbf{j}_{c}\,\mathbf{P}_k\,\mathbf{j}_{c}^\top + R}, \\
    \hat{\boldsymbol{\theta}}_{k+1} &= \hat{\boldsymbol{\theta}}_k + \mathbf{K}_k z_k, \\
    \mathbf{P}_{k+1} &= (\mathbf{I} - \mathbf{K}_k\,\mathbf{j}_{c})\,\mathbf{P}_k.
\end{align}

\subsection{Sample selection strategy}
To maximize test efficiency, the next code edge to be measured is selected based on the expected reduction in model uncertainty.
For each candidate code $c$, the scalar gain is computed as:

\begin{equation}
\label{eq:gain}
    \mathrm{gain}(c) = \frac{\mathbf{j}_c\,\mathbf{P}_k\,\mathbf{j}_c^\top}{\mathbf{j}_c\,\mathbf{P}_k\,\mathbf{j}_c^\top + R},
\end{equation}

which quantifies how much the current parameter uncertainty would be reduced by performing a measurement at code $c$.
The next sample location is chosen by maximizing this gain:

\begin{equation}
\label{eq:argmax}
    c^* = \arg\max_c \text{gain}(c).
\end{equation}

Since the Jacobian is precomputed and stored row-wise, this selection process can be efficiently implemented using parallel computation across all candidate codes.

\subsection{Measurement and shift computation}
\label{secsub:ProposedMethod_Measurement}
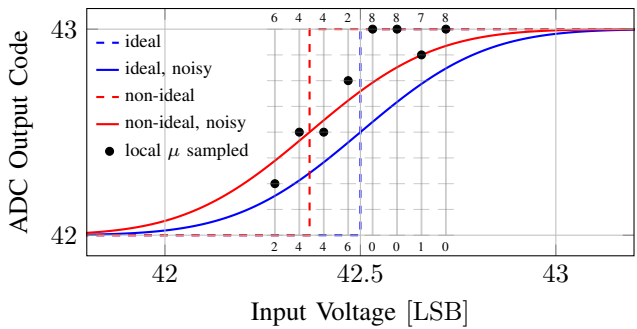
\begin{figure}[t]
    \centering
    \input{Plots/MiniSweep/MiniSweep}
    \caption{
    Example of a sweep used to estimate an ADC code edge under noise.
    The high-resolution DAC used for sampling here has $4$ bits more resolution than the ADC under test, enabling voltage steps of $1/16\,\mathrm{LSB}$.
    A total of 64 samples are uniformly distributed over a range of $\pm 0.25\,LSB$ around the predicted transition, resulting in 8 DAC input levels, each tested with 8 samples.
    Gray vertical lines indicate sampled input voltages, each with a vertical scale from 0 to 1 (in steps of $1/8$) showing how many samples hit code 42 or 43.
    }
    \label{fig:MiniSweep}
\end{figure}

Once a code $c^*$ is selected, the corresponding code edge position is estimated by applying a fine voltage sweep around the predicted transition point.
A representative example of the localized sweep is shown in \autoref{fig:MiniSweep}.
The input voltages are generated using a high-resolution DAC capable of producing steps finer than $1\,\mathrm{LSB}$ of the SAR ADC.
The sweep is centered around the current model estimate of the code edge, which initially aligns with the ideal edge location but gradually shifts as the parameter estimate is refined over successive iterations.
Each voltage in the sweep is applied to the ADC, and the resulting output code is recorded.

A batch of $M$ such measurements yields output codes $y_1, \dots, y_M$.
The average shift relative to the expected code $c^*$ is used as a proxy for the true deviation:

\begin{equation}
    z_k = \frac{1}{M} \sum_{i=1}^M y_i - c^* + 0.5.
\end{equation}

The offset of 0.5 centers the estimated edge between codes.
The sweep range is chosen based on the expected noise level to ensure adequate coverage.
If fewer DAC codes are available within this range than desired samples, inputs are distributed evenly across them, reusing DAC codes as needed.

\subsection{Covariance inflation based on unexpected innovation}
\label{secsub:ProposedMethod_Inflation}
While the EKF reduces the covariance matrix $\mathbf{P}_k$ as new measurements are incorporated, this reduction assumes that observed residuals are consistent with the predicted uncertainty.
When using the previously described measurement strategy based on a localized sweep around the predicted position of a code edge, unexpectedly large residuals $z_k$ can still occur, even when the filter's current covariance suggests high confidence.
These are caused by quantization effects, sampling noise, or uneven distribution of input voltages.

To address this, the proposed UGLMS includes a residual-based adaptive inflation strategy inspired by the approach presented by \cite{anderson2001ensemble}.
After each Kalman update, the measurement $z_k$ is compared to the predicted measurement variance $S_k$:

\begin{equation}
    S_k = \mathbf{j}_{c^*}\,\mathbf{P}_k\,\mathbf{j}_{c^*}^\top + R,
\end{equation}

and the Normalized Innovation Squared (NIS) is computed as:

\begin{equation}
    \mathrm{NIS}_k = \frac{z_k^2}{S_k}.
\end{equation}

If $\text{NIS}$ exceeds a user-defined threshold $\tau$, this is interpreted as a sign that the actual residual is larger than expected, indicating that the current model may not fully capture the system behavior.
The threshold is chosen empirically based on observed residual distributions, and can be adjusted to tune the sensitivity of the inflation mechanism.
In such cases, the covariance matrix is uniformly inflated:

\begin{equation}
    \mathbf{P}_{k+1} \leftarrow \alpha \cdot \mathbf{P}_{k+1}, \quad \text{for } \text{NIS} > \tau,
\end{equation}

where $\alpha > 1.0$ is the inflation factor.
This increases the estimated uncertainty and allows the next few measurements to exert more influence, preventing the filter from prematurely locking into a potentially incorrect solution.

\subsection{DNL \& INL estimation}
\label{secsub:ProposedMethod_INLEstimation}
After convergence, the final mismatch estimate $\hat{\boldsymbol{\theta}}$ is used to compute the full set of code edge positions $\mathrm{CE}[c] = f_c(\hat{\boldsymbol{\theta}})$ via the mismatch-aware model.
Having all $\mathrm{CE}[c]$ values directly allows reconstruction of the DNL and INL values according to the definitions given in \autoref{eq:DNL} and \autoref{eq:INL}.

\section{Experimental results}
\label{sec:Experimental_Results}
To evaluate the proposed UGLMS method, a series of simulation experiments were performed across different ADC resolutions, noise levels, and local sweep configurations.
The proposed algorithm was implemented in C for computational efficiency.
The SAR ADC model was parameterized using mismatch vectors derived from a dataset of real, measured ADCs.
To protect proprietary data, the original distributions were resampled and slightly shifted, while preserving the observed standard deviations and structural characteristics.

\subsection{Single run demonstration}
\begin{figure*}[t]
    \centering
    \input{Plots/SingleRun_Group/SingleRun_Group}
    \caption{
    INL and DNL of a 12-bit ADC reconstructed after 200 iterations using 64 samples per local sweep.
    Top: estimated and true INL/DNL. Bottom: absolute deviation. Measurement noise was set to $1.0\,\mathrm{LSB}$ RMS.
    }
    \label{fig:Group_INL_DNL}
\end{figure*}
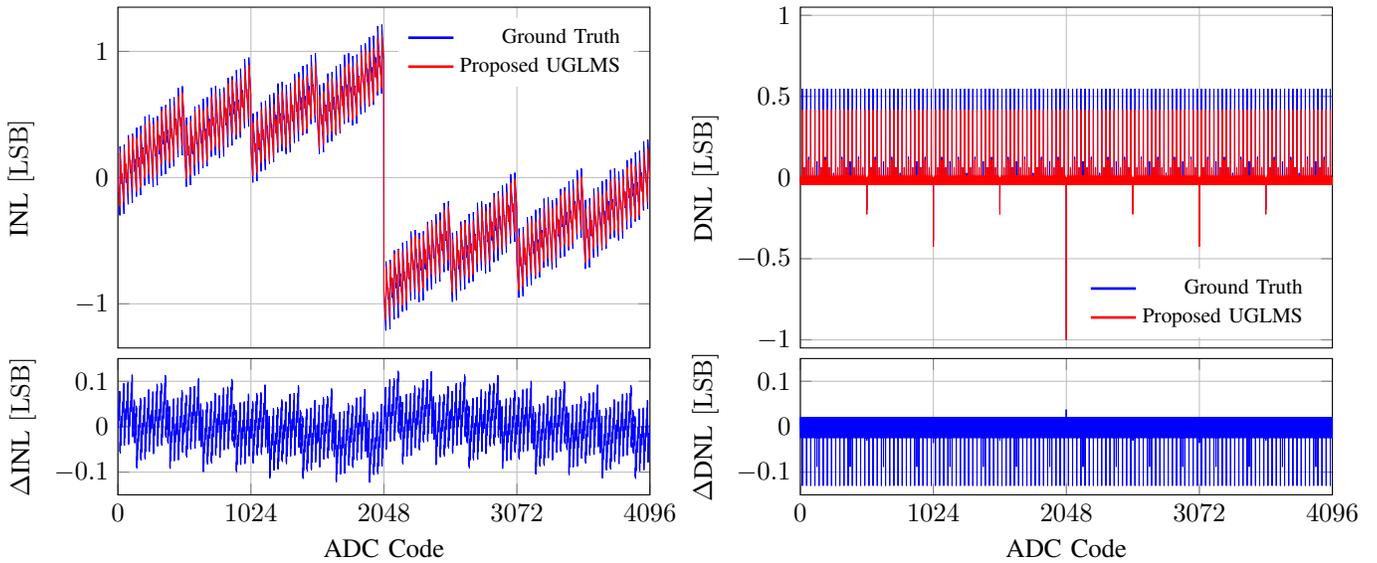

\autoref{fig:Group_INL_DNL} shows the reconstructed INL and DNL of a 12-bit ADC after 200 iterations.
The estimated INL curve closely follows the true INL, with absolute deviations remaining below $0.15\,\mathrm{LSB}$, which is well within common requirements.
The DNL results exhibit similarly high accuracy. Notably, a missing code, reflected as a sharp $-1\,\mathrm{LSB}$ drop in the true DNL, is correctly identified by the proposed method, demonstrating its robustness in capturing even discrete discontinuities.

\subsection{Sensitivity to test parameters}
\label{sec:Experimental_ConvergenceSweep}
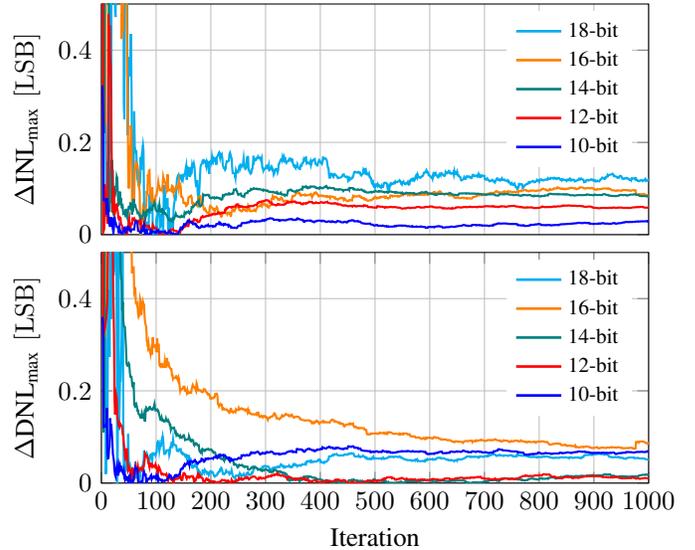
\begin{figure}[t]
    \input{Plots/VarIter_BitWidth/VarIter_BitWidth}
    \caption{
    Convergence of maximum INL and DNL estimation error over 1000 iterations for ADCs with different resolutions. 
    Measurement noise was $1.0\,\mathrm{LSB}$ RMS; each measurement sweep used 64 samples.
    }
    \label{fig:VarIter_BitWidth}
\end{figure}

\textbf{Resolution:} \autoref{fig:VarIter_BitWidth} shows convergence for ADCs from 10 to 18 bits.
As expected, higher resolutions require more iterations due to the larger parameter space and denser code edge distribution.
While lower resolutions generally converge faster, occasional non-monotonic behavior occurs - for instance, the 16-bit DNL error converges slightly slower than the 18-bit case.
This variation is likely caused by edge positions falling near the sweep boundary, noise-induced distortion, or quantization effects.
In such cases, a sweep may not fully capture the edge transition, reducing the quality of the update.
Increasing the sweep range could mitigate this, but at the cost of reduced sensitivity to fine corrections in later iterations.

\begin{figure}[t]
    \setlength{\abovecaptionskip}{-5pt} 
    \setlength{\belowcaptionskip}{-5pt} 
    \input{Plots/VarIter_BatchSampleCount/VarIter_BatchSampleCount}
    \caption{
    Convergence of maximum INL and DNL estimation error for different samples per measurement sweep.
    A 16-bit ADC in combination with $1.0\,LSB$ RMS noise was used.}
    \label{fig:VarIter_BatchSampleCount}
\end{figure}
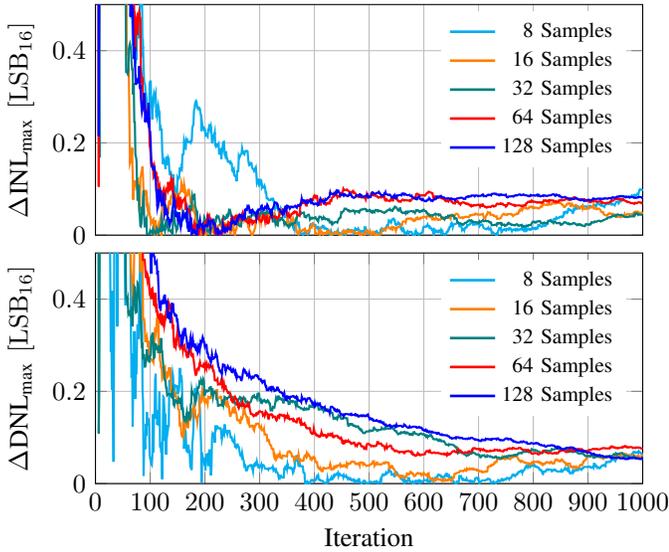

\textbf{Sweep sample count:} \autoref{fig:VarIter_BatchSampleCount} demonstrates that lower sweep sizes (e.g., 8 samples) lead to faster but noisier convergence.
In particular, the 8-sample configuration shows temporary divergence in INL near iteration 200 before recovering.
This is due to increased residual variance, which causes stronger Kalman updates and a higher likelihood of missteps.
Larger sample sweeps (e.g., 64 or 128 samples) suppress noise more effectively, leading to smoother convergence but slower initial progression.

\begin{figure}[t]
    \setlength{\abovecaptionskip}{-5pt} 
    \setlength{\belowcaptionskip}{-5pt} 
    \input{Plots/VarIter_Noise/VarIter_Noise}
    \caption{
    Convergence of maximum INL and DNL estimation error under different noise levels.
    A 16-bit ADC and 64 samples per measurement sweep were used.}
    \label{fig:VarIter_Noise}
\end{figure}
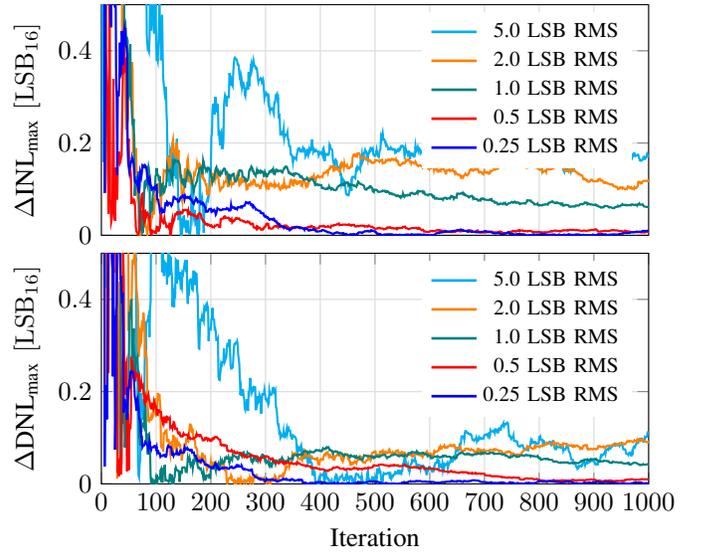

\textbf{Input noise:} \autoref{fig:VarIter_Noise} shows the effect of varying input noise levels from $0.25$ to $5.0\,\mathrm{LSB}$ RMS.
At low noise levels, convergence is fast and smooth.
As noise increases, the estimation error in both INL and DNL grows, and convergence becomes more gradual.
Still, the filter maintains functional stability even at high noise levels, with final estimation errors remaining within $\pm 0.2\,\mathrm{LSB}$.
This confirms the method's robustness under varying noise conditions.

All convergence plots exhibit a typical pattern: rapid early improvement followed by a slight increase in error during later iterations.
This is a known EKF artifact: without process noise, shrinking covariance leads to overconfidence, allowing small residuals to gradually shift the estimate.

\subsection{Statistical accuracy}
\label{secsub:ExperimentalResults_Statistics}
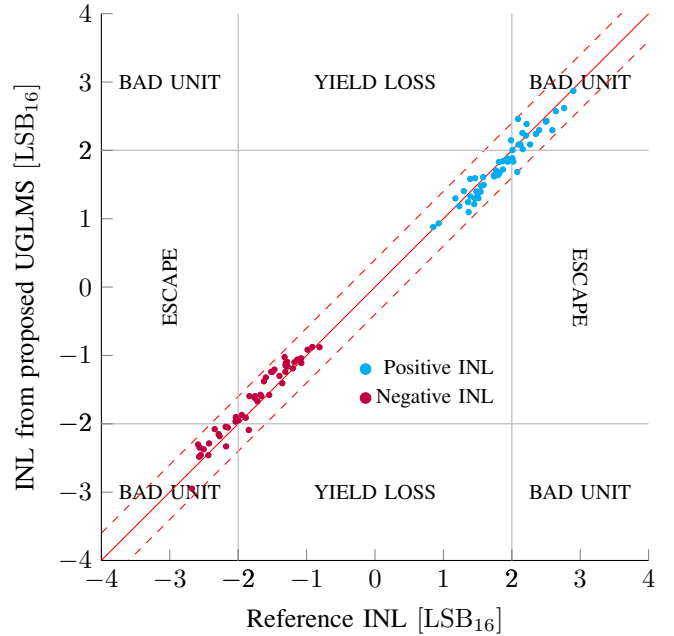
\begin{figure}[t]
    \setlength{\belowcaptionskip}{-5pt} 
    \input{Plots/StatisticalAnalysisINL/StatisticalAnalysisINL}
    \caption{
    INL correlation diagram showing the maximum INL value of 100 16-bit SAR ADCs tested using the UGLMS method.
    Each result is based on 200 iterations with 128 sweep samples and $1.0\,\mathrm{LSB}$ RMS input noise.
    }
    \label{fig:StatisticalDeviation}
\end{figure}

To evaluate robustness across device variation, a statistical analysis was performed on 100 synthetic 16-bit SAR ADCs.
Each device was assigned a unique mismatch vector sampled from a statistically perturbed version of the measured dataset and tested using 200 iterations with 128 sweep samples and $1.0\,\mathrm{LSB}$ RMS noise.
\autoref{fig:StatisticalDeviation} shows that all maximum INL deviations remain within $\pm 0.4\,\mathrm{LSB}$, with no bias or outliers.

\subsection{Test time}
Test time in UGLMS consists of two components: (1) measurement time for each selected code edge and (2) computation time to determine the next most informative code edge using the gain computation in \autoref{eq:gain} and \ref{eq:argmax}.
A breakdown of typical per-iteration and total test times for representative configurations is summarized in \autoref{tab:testTime}.

Assuming a $1\,\mathrm{MS/s}$ sampling rate, sweep acquisition is fast - even for 128 samples.
The main bottleneck is target selection, which involves matrix-vector operations that scale with ADC resolution.
To hide this latency, a concurrent execution strategy is used: gain computation runs in parallel with the measurement of the previously selected code edge.

In practice, the gain evaluation is highly parallelizable and is already implemented as a parallel routine.
Further speedup is achievable through increased compute resources for dedicated GPU acceleration.

Importantly, several matrices used in the estimation process, such as the Jacobian, can be precomputed once and reused across devices of the same type.
This enables fast sequential testing of multiple ADCs without incurring setup time per device.
As such, this startup overhead is excluded from the reported per-device test time.

Finally, since the mismatch parameters are fully estimated by the end of the test, only a small amount of result data needs to be transferred.
The INL can be computed directly and efficiently from the final mismatch vector, requiring no further post-processing.
Therefore, total test time is dominated by the iterative acquisition and code selection process, while data transfer and final computation are negligible in comparison.

\begin{table}[h]
    \centering
    \caption{
    Test time comparison for SAR ADCs with different resolutions.
    The code selection time refers to the computational effort of selecting the next code to test, as defined in \autoref{eq:gain} and \ref{eq:argmax}.
    All tests executed 200 iterations with input noise of $1.0\,\mathrm{LSB}$ RMS.
    }
    \label{tab:testTime}
    \begin{tabular}{l@{\hskip 6pt}|@{\hskip 6pt}c@{\hskip 6pt}c@{\hskip 6pt}c@{\hskip 6pt}c@{\hskip 6pt}c}
        \hline
        Resolution & 10-bit & 12-bit & 14-bit & 16-bit & 18-bit \\
        Samples/sweep & 64 & 64 & 128 & 128 & 128 \\
        \hline
        Per-code selection  & $88.0\mu \mathrm{s}$ & $89.5\mu \mathrm{s}$ & $141.0\mu \mathrm{s}$ & $300.5\mu \mathrm{s}$ & $1181.0\mu \mathrm{s}$ \\
        Selection total  & $17.6\,\mathrm{ms}$ & $17.9\,\mathrm{ms}$ & $28.2\,\mathrm{ms}$ & $60.1\,\mathrm{ms}$ & $236.2\,\mathrm{ms}$ \\
        Total          & $18.5\,\mathrm{ms}$ & $18.7\,\mathrm{ms}$ & $31.6\,\mathrm{ms}$ & $61.2\,\mathrm{ms}$ & $238.8\,\mathrm{ms}$ \\
        \hline
    \end{tabular}
\end{table}

\section{Discussion}
\label{sec:Discussion}
The proposed UGLMS method achieves a favorable trade-off between test accuracy, measurement efficiency, and real-time capability.
It consistently reconstructs INL and DNL within a sub-$0.4\,\mathrm{LSB}$ margin across all tested scenarios after only a few hundred iterations.
The statistical robustness of the method was further confirmed through a Monte Carlo analysis on 100 independently parameterized ADCs, each exhibiting realistic mismatch variations derived from measured chips.

In contrast to prior approaches such as uSMILE and its successors \cite{aswin2023new, chaganti2018fast}, which compress the number of required measurements through segmentation and offline model fitting, the UGLMS method emphasizes real-time estimation using lightweight recursive updates.
This removes the need for large-scale data collection and post-processing, making UGLMS more suitable for tightly integrated production test loops.

\begin{figure}[t]
    \setlength{\abovecaptionskip}{-5pt} 
    \setlength{\belowcaptionskip}{-5pt} 
    \input{Plots/HistComparison/HistComparison_INL}
    \caption{
    INL and DNL comparison for the 12-bit ADC shown in \autoref{fig:Group_INL_DNL}, reconstructed after 200 iterations using 64 samples per measurement sweep. 
    Results are compared to a full Ramp Histogram Test (RHT) with 128 hits per code.
    Measurement noise was set to $1.0\,\mathrm{LSB}$ RMS.
    }
    \label{fig:HistComparison}
\end{figure}
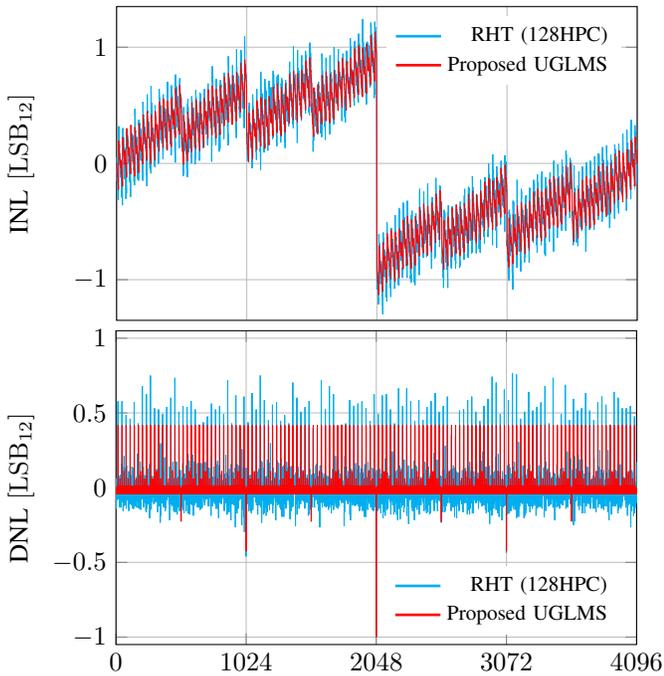

\autoref{fig:HistComparison} offers a direct visual comparison against a traditional 128 hits-per-code (HPC) ramp histogram test (RHT), which represents a generously sampled baseline.
Despite its high HPC, the histogram method shows visible statistical noise and fails to fully resolve small-scale structures, especially near sharp transitions.
In contrast, the proposed UGLMS provides a significantly smoother reconstruction with a fraction of the total samples and no post-processing overhead.

While conventional histogram tests benefit from their simplicity and long-standing qualification, they scale poorly with resolution and often require minutes of runtime or prohibitively large memory buffers.
In comparison, the adaptive UGLMS approach completes testing in tens of milliseconds (see \autoref{tab:testTime}) and directly provides mismatch parameters, enabling full INL reconstruction with minimal data transfer.

It is worth noting that the algorithm exceeds a one-HPC equivalent in total sample count.
However, this is largely due to the interleaved computation strategy that enables measurement sweeps to be conducted concurrently with code selection.
As such, runtime rather than raw sample count becomes the more meaningful optimization target.

Compared to existing model-driven approaches, the UGLMS method also outperforms in total test time. For instance, \cite{aswin2023new} reports test-time reductions to approximately $3\%$ of a one-HPC ramp test, yet relies on dense reconstruction from extrapolated samples and offline processing routines that alone take $890\,\mathrm{ms}$.
Similarly, uSMILE-based techniques \cite{chaganti2018fast} require model segmentation and matrix solving after acquisition, and estimate errors only after full INL/DNL reconstruction.
In contrast, our method performs both estimation and measurement inline, completing a 16-bit ADC test, including all computa-tion, in under $70\,\mathrm{ms}$, without additional reconstruction steps.

While the presented evaluation relies on a fixed number of iterations for fair comparison, UGLMS naturally supports adaptive termination.
Once the uncertainty encoded in the covariance matrix falls below a defined threshold, sampling can be halted, ensuring that measurements are taken only as long as they contribute meaningful information.

\section{Conclusion}
\label{sec:Conclusion}
This paper presents Uncertainty-Guided Live Measurement Sequencing (UGLMS), a closed-loop test strategy for SAR ADCs that estimates capacitor mismatches through targeted local measurements and real-time Kalman filtering.
Instead of collecting full-range datasets, UGLMS selects measurement points based on information gain and updates the mismatch model online.
The INL and DNL are reconstructed from the estimated parameters without requiring post-processing.

Compared to conventional histogram methods and recent model-based approaches, UGLMS shifts focus from pure sample minimization toward overall test efficiency.
Sample counts per code may exceed 1 HPC, but due to interleaved measurement and computation, overall test time remains low, with mismatch estimates available immediately at test end.
For typical 16-bit configurations, test durations under $70\,\mathrm{ms}$ have been demonstrated, with sub-$0.4\,\mathrm{LSB}$ accuracy.

Recent methods such as USER-SMILE and curve-fitting based INL recovery achieve high data compression by reducing the number of required samples \cite{aswin2023new, chaganti2018low, chaganti2018fast, chen2015ultrafast}.
However, they typically rely on post-processing to reconstruct linearity metrics.
In contrast, UGLMS performs all estimation steps during test execution, enabling faster test cycles and immediate availability of mismatch parameters for downstream analysis.
Moreover, the framework supports adaptive termination based on model uncertainty, allowing tests to conclude automatically once a target confidence level is reached.

\section*{Acknowledgements}
This research was supported by Advantest as part of the Graduate School “Intelligent Methods for Test and Reliability” (GS-IMTR) at University of Stuttgart.

\section*{Declaration}
During the preparation of this work, the authors used ChatGPT 4o and Grammarly in order to polish parts of text for better readability. After using these tools/services, the authors reviewed and edited the content as needed and take full responsibility for the content of the publication.
\newpage

\IEEEtriggeratref{11}
\bibliographystyle{IEEEtran}
\bibliography{literature}

\end{document}

%% file: Plots/MiniSweep/MiniSweep.tex
\begin{tikzpicture}
\begin{axis}[
    width=\linewidth,
    height=0.55\linewidth,
    xlabel={Input Voltage $[\mathrm{LSB}]$},
    ylabel={ADC Output Code},
    ymin=41.9, ymax=43.1,
    xmin=41.8, xmax=43.2,
    xtick distance=0.5,
    ytick={42, 43},
    grid=major,
    axis on top,
    legend style={
        font=\scriptsize,           
        draw=none,                  
        fill=none,                  
        at={(0.0, 0.94)},           
        anchor=north west,
        cells={align=left},          
        legend image code/.code={
        \draw[##1] (0cm,0cm) -- (0.3cm,0cm);  
        }
    },
    legend cell align={left}
]

\addplot[thick, dashed, blue] table {Plots/MiniSweep/step_ideal.dat};
\addlegendentry{ideal}

\addplot[blue, thick] table {Plots/MiniSweep/prob_ideal.dat};
\addlegendentry{ideal, noisy}

\addplot[thick, dashed, red] table {Plots/MiniSweep/step_shifted.dat};
\addlegendentry{non-ideal}

\addplot[red, thick] table {Plots/MiniSweep/prob_shifted.dat};
\addlegendentry{non-ideal, noisy}

\addplot[
    only marks,
    mark=*,
    mark size=1.5pt,
    black,
    legend image code/.code={
        \draw[mark=*, mark options={fill=black}] plot coordinates {(-0.19,0)};
    }
] table {Plots/MiniSweep/lollipop_means.dat};
\addlegendentry{local $\mu$ sampled}

\addplot[
    gray,
    opacity=0.4,
    line width=0.7pt
] table [x index=0, y index=1] {Plots/MiniSweep/lollipop_stems.dat};

\addplot[
    only marks,
    mark=-,
    mark size=3pt,
    gray,
    opacity=0.4,
] table {Plots/MiniSweep/lollipop_ticks.dat};

\node[anchor=north, font=\tiny] at (axis cs:42.28, 42.01) {2};
\node[anchor=south, font=\tiny] at (axis cs:42.28, 42.99) {6};

\node[anchor=north, font=\tiny] at (axis cs:42.3425, 42.01) {4};
\node[anchor=south, font=\tiny] at (axis cs:42.3425, 42.99) {4};

\node[anchor=north, font=\tiny] at (axis cs:42.405, 42.01) {4};
\node[anchor=south, font=\tiny] at (axis cs:42.405, 42.99) {4};

\node[anchor=north, font=\tiny] at (axis cs:42.4675, 42.01) {6};
\node[anchor=south, font=\tiny] at (axis cs:42.4675, 42.99) {2};

\node[anchor=north, font=\tiny] at (axis cs:42.53, 42.01) {0};
\node[anchor=south, font=\tiny] at (axis cs:42.53, 42.99) {8};

\node[anchor=north, font=\tiny] at (axis cs:42.5925, 42.01) {0};
\node[anchor=south, font=\tiny] at (axis cs:42.5925, 42.99) {8};

\node[anchor=north, font=\tiny] at (axis cs:42.655, 42.01) {1};
\node[anchor=south, font=\tiny] at (axis cs:42.655, 42.99) {7};

\node[anchor=north, font=\tiny] at (axis cs:42.7175, 42.01) {0};
\node[anchor=south, font=\tiny] at (axis cs:42.7175, 42.99) {8};

\end{axis}
\end{tikzpicture}

%% file: Plots/SingleRun_Group/SingleRun_Group.tex
\begin{tikzpicture}
\begin{groupplot}[
    group style={
        group size=2 by 2,
        vertical sep=4pt,
        horizontal sep=2cm,
    },
    width=0.39\linewidth,
    height=0.25\linewidth,
    scale only axis=true,
    xmin=0, xmax=4096,
    grid=major,
    legend style={draw=none, font=\footnotesize},
    legend cell align={right},
    legend image post style={line width=1pt},
]

\nextgroupplot[
    ylabel={INL $[\mathrm{LSB}]$},
    ylabel style={yshift=8.5pt},
    ymin=-1.35, ymax=1.35,
    xtick distance=1024,
    xmajorticks=false,
    legend pos=north east,
]
\addplot[blue] table {Plots/SingleRun_INL/true_inl.dat};
\addlegendentry{Ground Truth};
\addplot[red] table {Plots/SingleRun_INL/est_inl.dat};
\addlegendentry{Proposed UGLMS};

\nextgroupplot[
    ylabel={DNL $[\mathrm{LSB}]$},
    ymin=-1.05, ymax=1.05,
    xtick distance=1024,
    xmajorticks=false,
    legend pos=south east,
]
\addplot[blue] table {Plots/SingleRun_DNL/true_dnl.dat};
\addlegendentry{Ground Truth};
\addplot[red] table {Plots/SingleRun_DNL/est_dnl.dat};
\addlegendentry{Proposed UGLMS};

\nextgroupplot[
    xlabel={ADC Code},
    ylabel={$ \Delta $INL $[\mathrm{LSB}]$},
    ymin=-0.15, ymax=0.15,
    xtick distance=1024,
    xticklabel style={/pgf/number format/1000 sep=},
    ytick distance=0.1,
    height=0.1\linewidth,
]
\addplot[blue] table {Plots/SingleRun_INL/diff_inl.dat};

\nextgroupplot[
    xlabel={ADC Code},
    ylabel={$ \Delta $DNL $[\mathrm{LSB}]$},
    ymin=-0.15, ymax=0.15,
    xtick distance=1024,
    xticklabel style={/pgf/number format/1000 sep=},
    ytick distance=0.1,
    height=0.1\linewidth,
]
\addplot[blue] table {Plots/SingleRun_DNL/diff_dnl.dat};

\end{groupplot}

\end{tikzpicture}

%% file: Plots/VarIter_BitWidth/VarIter_BitWidth.tex
\begin{tikzpicture}
\begin{axis}[
    ylabel={$\Delta$INL\textsubscript{max} $[\mathrm{LSB}]$},
    grid=major,
    ymin=0,
    ymax=0.5,
    xmin=0,
    xmax=1000,
    xtick distance=100,
    xticklabel style={/pgf/number format/1000 sep=},
    xmajorticks=false,
    width=\linewidth,
    height=0.525\linewidth,
    legend pos=north east,
    legend cell align={left},
    legend style={draw=none, font=\footnotesize},
    legend image post style={line width=1pt}, 
]
\addplot[cyan, thick] table {Plots/VarIter_BitWidth/inl_deviation_N18.dat};
\addlegendentry{18-bit};
\addplot[orange, thick] table {Plots/VarIter_BitWidth/inl_deviation_N16.dat};
\addlegendentry{16-bit};
\addplot[teal, thick] table {Plots/VarIter_BitWidth/inl_deviation_N14.dat};
\addlegendentry{14-bit};
\addplot[red, thick] table {Plots/VarIter_BitWidth/inl_deviation_N12.dat};
\addlegendentry{12-bit};
\addplot[blue, thick] table {Plots/VarIter_BitWidth/inl_deviation_N10.dat};
\addlegendentry{10-bit};
        
\end{axis}
\end{tikzpicture}

\begin{tikzpicture}
\begin{axis}[
    xlabel={Iteration},
    ylabel={$\Delta$DNL\textsubscript{max} $[\mathrm{LSB}]$},
    grid=major,
    ymin=0,
    ymax=0.5,
    xmin=0,
    xmax=1000,
    xtick distance=100,
    xticklabel style={/pgf/number format/1000 sep=},
    width=\linewidth,
    height=0.525\linewidth,
    legend pos=north east,
    legend cell align={left},
    legend style={draw=none, font=\footnotesize},
    legend image post style={line width=1pt}, 
]
\addplot[cyan, thick] table {Plots/VarIter_BitWidth/dnl_deviation_N18.dat};
\addlegendentry{18-bit};
\addplot[orange, thick] table {Plots/VarIter_BitWidth/dnl_deviation_N16.dat};
\addlegendentry{16-bit};
\addplot[teal, thick] table {Plots/VarIter_BitWidth/dnl_deviation_N14.dat};
\addlegendentry{14-bit};
\addplot[red, thick] table {Plots/VarIter_BitWidth/dnl_deviation_N12.dat};
\addlegendentry{12-bit};
\addplot[blue, thick] table {Plots/VarIter_BitWidth/dnl_deviation_N10.dat};
\addlegendentry{10-bit};
        
\end{axis}
\end{tikzpicture}

%% file: Plots/VarIter_BatchSampleCount/VarIter_BatchSampleCount.tex
\begin{tikzpicture}
\begin{axis}[
    ylabel={$\Delta$INL\textsubscript{max} $[\mathrm{LSB_{16}}]$},
    grid=major,
    ymin=0,
    ymax=0.5,
    xmin=0,
    xmax=1000,
    xmajorticks=false, 
    xtick distance=100,
    xticklabel style={/pgf/number format/1000 sep=},
    width=\linewidth,
    height=0.525\linewidth,
    legend pos=north east,
    legend cell align={right},
    legend style={draw=none, font=\footnotesize},
    legend image post style={line width=1pt}, 
]
\addplot[cyan, thick] table {Plots/VarIter_BatchSampleCount/inl_deviation_B8.dat};
\addlegendentry{8 Samples};
\addplot[orange, thick] table {Plots/VarIter_BatchSampleCount/inl_deviation_B16.dat};
\addlegendentry{16 Samples};
\addplot[teal, thick] table {Plots/VarIter_BatchSampleCount/inl_deviation_B32.dat};
\addlegendentry{32 Samples};
\addplot[red, thick] table {Plots/VarIter_BatchSampleCount/inl_deviation_B64.dat};
\addlegendentry{64 Samples};
\addplot[blue, thick] table {Plots/VarIter_BatchSampleCount/inl_deviation_B128.dat};
\addlegendentry{128 Samples};
        
\end{axis}
\end{tikzpicture}

\begin{tikzpicture}
\begin{axis}[
    xlabel={Iteration},
    ylabel={$\Delta$DNL\textsubscript{max} $[\mathrm{LSB_{16}}]$},
    grid=major,
    ymin=0,
    ymax=0.5,
    xmin=0,
    xmax=1000,
    xtick distance=100,
    xticklabel style={/pgf/number format/1000 sep=},
    width=\linewidth,
    height=0.525\linewidth,
    legend pos=north east,
    legend cell align={right},
    legend style={draw=none, font=\footnotesize},
    legend image post style={line width=1pt}, 
]
\addplot[cyan, thick] table {Plots/VarIter_BatchSampleCount/dnl_deviation_B8.dat};
\addlegendentry{8 Samples};
\addplot[orange, thick] table {Plots/VarIter_BatchSampleCount/dnl_deviation_B16.dat};
\addlegendentry{16 Samples};
\addplot[teal, thick] table {Plots/VarIter_BatchSampleCount/dnl_deviation_B32.dat};
\addlegendentry{32 Samples};
\addplot[red, thick] table {Plots/VarIter_BatchSampleCount/dnl_deviation_B64.dat};
\addlegendentry{64 Samples};
\addplot[blue, thick] table {Plots/VarIter_BatchSampleCount/dnl_deviation_B128.dat};
\addlegendentry{128 Samples};
        
\end{axis}
\end{tikzpicture}

%% file: Plots/VarIter_Noise/VarIter_Noise.tex
\begin{tikzpicture}
\begin{axis}[
    ylabel={$\Delta$INL\textsubscript{max} $[\mathrm{LSB_{16}}]$},
    grid=major,
    grid style={gray!25, line width=0.5pt},
    ymin=0,
    ymax=0.5,
    xmin=0,
    xmax=1000,
    xmajorticks=false, 
    xtick distance=100,
    xticklabel style={/pgf/number format/1000 sep=},
    width=\linewidth,
    height=0.525\linewidth,
    legend pos=north east,
    legend cell align={right},
    legend style={draw=none, font=\footnotesize},
    legend image post style={line width=1pt}, 
]
\addplot[cyan, thick] table {Plots/VarIter_Noise/inl_deviation_S500.dat};
\addlegendentry{5.0 LSB RMS};
\addplot[orange, thick] table {Plots/VarIter_Noise/inl_deviation_S200.dat};
\addlegendentry{2.0 LSB RMS};
\addplot[teal, thick] table {Plots/VarIter_Noise/inl_deviation_S100.dat};
\addlegendentry{1.0 LSB RMS};
\addplot[red, thick] table {Plots/VarIter_Noise/inl_deviation_S50.dat};
\addlegendentry{0.5 LSB RMS};
\addplot[blue, thick] table {Plots/VarIter_Noise/inl_deviation_S25.dat};
\addlegendentry{0.25 LSB RMS};
        
\end{axis}
\end{tikzpicture}

\begin{tikzpicture}
\begin{axis}[
    xlabel={Iteration},
    ylabel={$\Delta$DNL\textsubscript{max} $[\mathrm{LSB_{16}}]$},
    grid=major,
    grid style={gray!25, line width=0.5pt},
    ymin=0,
    ymax=0.5,
    xmin=0,
    xmax=1000,
    xtick distance=100,
    xticklabel style={/pgf/number format/1000 sep=},
    width=\linewidth,
    height=0.525\linewidth,
    legend pos=north east,
    legend cell align={right},
    legend style={draw=none, font=\footnotesize},
    legend image post style={line width=1pt}, 
]
\addplot[cyan, thick] table {Plots/VarIter_Noise/dnl_deviation_S500.dat};
\addlegendentry{5.0 LSB RMS};
\addplot[orange, thick] table {Plots/VarIter_Noise/dnl_deviation_S200.dat};
\addlegendentry{2.0 LSB RMS};
\addplot[teal, thick] table {Plots/VarIter_Noise/dnl_deviation_S100.dat};
\addlegendentry{1.0 LSB RMS};
\addplot[red, thick] table {Plots/VarIter_Noise/dnl_deviation_S50.dat};
\addlegendentry{0.5 LSB RMS};
\addplot[blue, thick] table {Plots/VarIter_Noise/dnl_deviation_S25.dat};
\addlegendentry{0.25 LSB RMS};
        
\end{axis}
\end{tikzpicture}

%% file: Plots/StatisticalAnalysisINL/StatisticalAnalysisINL.tex
\begin{tikzpicture}
\begin{axis}[
    width=\linewidth,
    height=\linewidth,
    axis lines=left,
    axis line style={-},
    xmin=-4, xmax=4,
    ymin=-4, ymax=4,
    restrict x to domain=-4.5:4.5,
    restrict y to domain=-4.5:4.5,
    xlabel={{Reference INL $[\mathrm{LSB_{16}}]$}},
    ylabel={{INL from proposed UGLMS $[\mathrm{LSB_{16}}]$}},
    xtick distance=1,
    ytick distance=1,
    extra x ticks={-2,2},
    extra y ticks={-2,2},
    extra x tick style={grid=major},
    extra y tick style={grid=major},
    grid style={gray!50, line width=0.5pt},
    tick align=inside,
    axis on top,
    legend cell align={right},
    legend style={at={(0.6, 0.26)}, anchor=south, draw=none, fill=none, font=\footnotesize},
    legend image post style={mark size=2pt},
]

\node at (axis cs:0,3) [anchor=center, rotate=0] {\footnotesize{YIELD LOSS}};
\node at (axis cs:0,-3) [anchor=center, rotate=0] {\footnotesize{YIELD LOSS}};
\node at (axis cs:3,0) [anchor=center, rotate=270] {\footnotesize{ESCAPE}};
\node at (axis cs:-3,0) [anchor=center, rotate=90] {\footnotesize{ESCAPE}};
\node at (axis cs:3,3) [anchor=center, rotate=0] {\footnotesize{BAD UNIT}};
\node at (axis cs:3,-3) [anchor=center, rotate=0] {\footnotesize{BAD UNIT}};
\node at (axis cs:-3,3) [anchor=center, rotate=0] {\footnotesize{BAD UNIT}};
\node at (axis cs:-3,-3) [anchor=center, rotate=0] {\footnotesize{BAD UNIT}};

\addplot[domain=-4:4, red, forget plot] {x};

\addplot[domain=-4:4, red, dashed, forget plot] {x + 0.4};

\addplot[domain=-4:4, red, dashed, forget plot] {x - 0.4};

\addplot[
    only marks,
    mark=*,
    mark size=1pt,
    cyan,
] table [x index=0, y index=1, col sep=space]
{Plots/StatisticalAnalysisINL/inl_statistic_pos.dat};
\addlegendentry{Positive INL}

\addplot[
    only marks,
    mark=*,
    mark size=1pt,
    purple,
] table [x index=0, y index=1, col sep=space]
{Plots/StatisticalAnalysisINL/inl_statistic_neg.dat};
\addlegendentry{Negative INL}

\end{axis}
\end{tikzpicture}

%% file: Plots/HistComparison/HistComparison_INL.tex
\begin{tikzpicture}
\begin{axis}[
    ylabel={INL $[\mathrm{LSB_{12}}]$},
    ylabel style={yshift=8pt},
    grid=major,
    ymin=-1.35,
    ymax=1.35,
    xmin=0,
    xmax=4096,
    xmajorticks=false, 
    xtick distance=1024,
    width=0.96\linewidth,
    height=0.65\linewidth,
    legend pos=north east,
    legend cell align={right},
    legend style={draw=none, font=\footnotesize},
    legend image post style={line width=1pt}, 
]
\addplot[cyan] table {Plots/HistComparison/hist_est_inl.dat};
\addlegendentry{RHT (128HPC)};
\addplot[red] table {Plots/HistComparison/est_inl.dat};
\addlegendentry{Proposed UGLMS};
\end{axis}
\end{tikzpicture}

\begin{tikzpicture}
\begin{axis}[
    ylabel={DNL $[\mathrm{LSB_{12}}]$},
    grid=major,
    ymin=-1.05,
    ymax=1.05,
    xmin=0,
    xmax=4096,
    xtick distance=1024,
    xticklabel style={/pgf/number format/1000 sep=},
    width=0.96\linewidth,
    height=0.65\linewidth,
    legend pos=south east,
    legend cell align={right},
    legend style={draw=none, font=\footnotesize},
    legend image post style={line width=1pt}, 
]
\addplot[cyan] table {Plots/HistComparison/hist_est_dnl.dat};
\addlegendentry{RHT (128HPC)};
\addplot[red] table {Plots/HistComparison/est_dnl.dat};
\addlegendentry{Proposed UGLMS};
\end{axis}
\end{tikzpicture}